\documentclass[1p]{elsarticle}
\usepackage{graphicx} \usepackage{amsmath} \usepackage{times}
\usepackage{amssymb} \usepackage{mathrsfs} \usepackage{mathenv}
\usepackage{chemarr} \usepackage{color}
\definecolor{linkcolor}{rgb}{0,0,0.6} \usepackage[
  pdftex,colorlinks=true, pdfstartview=FitV, linkcolor= linkcolor,
  citecolor= linkcolor, urlcolor= linkcolor, hyperindex=true,
  hyperfigures=false] {hyperref}
	
\usepackage{url}
\usepackage{version}

\newcommand{\dd}{\text{d}}

\newcommand{\ee}{\text{e}}

\newcommand{\p}{\partial}

\newcommand{\kb}{k_\text{\tiny B}}

\newcommand{\g}{\gamma}

\newcommand{\fp}{f_\text{\tiny P}}

\newcommand{\alp}{\alpha}

\newcommand{\Fe}{F_\text{ext}}
\newcommand{\db}{\zeta_\text{\tiny B}}
\newcommand{\Fs}{F_\text{\tiny S}}
\newcommand{\Fb}{F_\text{\tiny B}}

\newcommand{\cA}{\mathcal{S}}
\newcommand{\cP}{\mathcal{P}}
\newcommand{\mf}{m^*}
\newcommand{\ci}{\chi_\text{in}}
\newcommand{\ce}{\chi_\text{ext}}
\newcommand{\cs}{\chi_\text{\tiny S}}
\newcommand{\cb}{\chi_\text{\tiny B}}

\def\GG{{\mathcal G}}

\providecommand{\avg}[1]{\left \langle #1 \right \rangle}
\providecommand{\pnt}[1]{\left ( #1 \right)}
\providecommand{\brt}[1]{\left [ #1 \right]}

\providecommand{\abs}[1]{\left | #1 \right|}
\providecommand{\f}[2]{\frac{ #1}{#2}}

\def\s{s}  
\newcommand\DXtilde{\stackrel{\sim}{\smash{\avg{\Delta x^2}}\rule{0pt}{1.1ex}}\hspace*{-0.8mm}}

\begin{document}
\begin{frontmatter}
\title{Generalized Langevin Equation with Hydrodynamic Backflow:
  Equilibrium Properties}

\author[msc]{\'Etienne Fodor}
  \ead{etienne.fodor@univ-paris-diderot.fr}

\author[lpmc]{Denis~S.~Grebenkov}
  \ead{denis.grebenkov@polytechnique.edu}

\author[msc]{Paolo Visco}
  \ead{paolo.visco@univ-paris-diderot.fr}

\author[msc,ucb]{Fr\'ed\'eric van Wijland}
  \ead{fvw@univ-paris-diderot.fr}

\address[msc]{
Laboratoire Mati\`ere et Syst\`emes Complexes, CNRS UMR 7057, \\
Universit\'e Paris Diderot, 10 rue Alice Domon et L\'eonie Duquet, 75205 Paris cedex 13, France}

\address[lpmc]{
Laboratoire de Physique de la Mati\`{e}re Condens\'{e}e (UMR 7643), \\ 
CNRS -- \'Ecole Polytechnique, 91128 Palaiseau, France}

\address[ucb]{Department of Chemistry, University of California, Berkeley, CA, 94720, USA}

\date{\today}

\begin{abstract}
We review equilibrium properties for the dynamics of a single 
particle evolving in a visco--elastic medium under the effect 
of hydrodynamic backflow which includes added mass and Basset 
force. Arbitrary equilibrium forces acting upon the particle 
are also included. We discuss the derivation of the explicit 
expression for the thermal noise correlation function that is 
consistent with the fluctuation-dissipation theorem. We rely
on general time-reversal arguments that apply irrespective of 
the external potential acting on the particle, but also allow 
one to retrieve existing results derived for free particles and particles in a harmonic trap. 
Some consequences for the analysis and interpretation of 
single-particle tracking experiments are briefly discussed.
\end{abstract}



\end{frontmatter}

\section{Introduction}

Single-particle tracking experiments can access dynamical, structural
and microrheological properties of complex visco-elastic media such as
polymer gels or living cells \cite{Rohrbach04,Wirtz09}.  Random
displacements of a tracer are often analyzed with the help of a
generalized Langevin equation which incorporates all relevant
interactions of the tracer, e.g., viscous or visco-elastic Stokes
force, inertial and hydrodynamic effects, active pulling by motor
proteins, and eventual optical trapping
\cite{Clercx92,Desposito09,Bruno09,Grebenkov11,Indei12a,Indei12b,Grebenkov13,Grebenkov14}.
Since several different mechanisms interplay in a complex medium, the
correct formulation of the underlying phenomenological model can be
sophisticated.  For instance, the correlation function of the thermal
noise has to be related, at equilibrium, to the memory kernels of the
generalized Stokes and Basset forces according to the
fluctuation-dissipation theorem. {A recent experiment by Kheifets {\it
    et al.}~\cite{Raizen14} tracking micrometer-sized glass beads in
  water or acetone reveals that equipartition is broken in equilibrium
  by a contribution involving the mass of the displaced fluid. This
  raises the question of which ingredients relating to the surrounding
  fluid will appear in other manifestations of equilibrium, such as
  the fluctuation-dissipation theorem.} 

In this paper, we investigate the equilibrium properties of a
generalized Langevin equation with hydrodynamic interactions and we
provide the correct noise correlation function, consistent with the
fluctuation-dissipation theorem. The role of the acceleration of the
displaced fluid is sorted out, thus justifying the assumption made in
\cite{Komura93} and amending that of
\cite{Grebenkov13,Grebenkov14}. Our analysis goes along the lines of
that of Baiesi {\it et al.}~\cite{maesbaiesiwynants}. Some
consequences for the analysis and interpretation of single-particle
tracking experiments are briefly discussed.

\section{Model}\label{sec:mod}

We are interested in the short time-scale motion of a tracer with mass
$m$ the displacement of which takes place in a complex visco-elastic
medium, such as a gel. For simplicity, we restrict here to the one
dimensional case, although generalization to two and three dimensions
is straightforward. We denote by $x(t)$ the tracer's position, and we
assume the tracer is subjected to an external force $\Fe$ and we
further allow ourselves the possibility to apply a small perturbation
force $\fp$. Newton's equation for the tracer reads
\begin{equation}\label{eq:dyn}
m \ddot x = \Fs + \Fb + \Fe + \fp + \xi
\,\,,
\end{equation}
where $\ddot x$ is the tracer's acceleration. In Eq. \eqref{eq:dyn},
in addition to the deterministic forces $\Fe$ and $\fp$, we have
included a Gaussian colored noise $\xi(t)$ accounting for the
interaction of the tracer with the heat bath. We have also included a
generalized Stokes force $\Fs$, which expresses the viscous friction
exerted by the fluid on the tracer. The latter force, when
coarse-graining out the degrees of freedom of the surrounding medium,
can be cast in the form \cite{Mori65,Hess83}
\begin{equation}\label{genStokes}
\Fs(t) = - \int\limits_{t_0}^\infty \dd t' \g(t-t') \dot x(t') ,
\end{equation}
where the memory kernel $\g(\tau)$ is causal (i.e., $\g(\tau) = 0$ for
$\tau < 0$), and the starting time $t_0$ is typically set either to
$-\infty$ or to $0$. A number of
experiments~\cite{Wilhelm,GalletSM,Gallet06,Bertseva12} in living
cells or in synthetic polymer solutions point to $\g$ being accurately
described by a power law \cite{Desposito09,Grebenkov11}, thereby
expressing that a hierarchy of time-scales is involved in viscous
friction for these complex media. Much less studied in a visco-elastic
medium is the Basset force $\Fb$ which we have also included in
Eq.~\eqref{eq:dyn} following \cite{Grebenkov13,Grebenkov14}. As much
as the usual inertia contribution $m\ddot{x}$, the Basset force in
usually negligible at the macroscopic observation time scales
considered in standard tracking experiments, but its effects have been
shown to be prominent at short time-scales in
\cite{Grimm11,Huang11,Indei12a,Indei12b,Grebenkov13,Grebenkov14}.
This force is related to the inertia of the boundary layer surrounding
the tracer. While the initial derivation for the expression of the
Basset force in terms of the tracer's position dates back to
Boussinesq for Newtonian fluids, Zwanzig and
Bixon~\cite{Zwanzig70,Schieber20133} provided a derivation of that
force for a visco-elastic fluid characterized by a memory kernel
$\gamma$ as in Eq.~\eqref{genStokes}. The generalized Basset force
then reads
\begin{equation}
\Fb(t) =- \f{m_\text{f}}{2} \ddot x(t) - \int\limits_{t_0}^\infty \dd
t' \db(t-t') \ddot x(t') ,
\end{equation}
where $m_\text{f}$ is the mass of the fluid displaced by the
tracer. The memory kernel $\db$ is causal as well, and can be argued
to be related to $\gamma$ in the following fashion
\begin{equation}
\label{eq:kernw}
\hat{\db}(\omega)=
3\sqrt{\f{m_\text{f}\hat{\g}(\omega)}{2i\omega}},\qquad
\tilde{\zeta}_\text{\tiny B}(\s) = 3\sqrt{\f{m_\text{f}\tilde{\g}(\s)}{2\s}} .
\end{equation}
where the hat and the tilde stand for the Fourier and the Laplace
transforms, respectively. In order to arrive at Eq.~\eqref{eq:kernw},
the argument put forward in \cite{Zwanzig70} goes as follows: for a
Newtonian fluid, one has $\hat{\zeta}_\text{\tiny B}(\omega)=6\pi
a^2\sqrt{\frac{\rho_\text{f}\eta}{i\omega}}$, where $a$ is the
tracer's radius. For a visco--elastic medium, the viscosity is to be
replaced with its frequency-dependent expression $\hat{\eta}(\omega)$,
thus leading to $\hat{\zeta}_\text{\tiny B}(\omega)=6\pi
a^2\sqrt{\frac{\rho_\text{f}\hat{\eta}}{i\omega}}$. Finally, with the
generalized Stokes law $\hat{\gamma}=6\pi\hat{\eta} a$ for spherical
tracers, we obtain Eq.~\eqref{eq:kernw}.  Note that the following
derivation does not rely on relation \eqref{eq:kernw} between memory
kernels $\g(t)$ and $\db(t)$, and it is thus valid in a more general
situation.

The question we now ask regards to thermal noise correlations
$\sigma(t-t')=\langle \xi(t)\xi(t')\rangle$ that we must impose to
ensure that in the absence of a perturbing force ($\fp=0$) and for a
conservative external force $\Fe$ that derives from a potential, the
tracer undergoes equilibrium and reversible dynamics, in agreement
with, {\it e.g.}, the fluctuation-dissipation theorem. {In the absence of the Basset force, this issue has been settled in the seminal paper by Kubo~\cite{0034-4885-29-1-306} and further discussed in the nice reviews by Mainardi {\it et al.}~\cite{mainardimpsp} or by H\"anggi~\cite{hanggireview}.} We begin by
recalling the expression of the fluctuation-dissipation theorem.

\section{Stating the Fluctuation--dissipation theorem}

The response of a position-dependent observable $A$ to an
infinitesimal external {perturbation} $\fp(t')$ is denoted by $\chi$ and it is defined by
\begin{equation}\label{eq:respd}
\chi(t,t') = \left.\f{\delta \avg{A(t)}}{\delta \fp(t')}\right|_{\fp=0}
\,\,.
\end{equation}
{Equilibrium first requires stationarity, namely time-translation invariance, so that
$\chi(t,t')=\chi(t-t')$ in the regime of interest.}
Causality ensures the response function vanishes if the measurement is
performed before the perturbation, when $t\leq t'$. The
fluctuation--dissipation theorem (FDT) states that in equilibrium the
response is related to the correlation between the observable and the
perturbation as~\cite{Callen}:
\begin{equation}\label{eq:fdtg}
\chi(t-t')=\beta \f{\p\avg{A(t)x(t')}}{\p t'} \Theta(t-t')
\,\,,
\end{equation}
where $\beta=1/(\kb T)$, $T$ is the bath temperature, and $\Theta$
denotes the Heaviside function.  {Stationarity also leads to $\avg{A(t)x(t')}=\avg{A(t-t')x(0)}$}.  The
FDT can be written without enforcing explicit causality as
\begin{equation}\label{eq:fdt}
\chi(\tau)-\chi(-\tau)=-\beta \frac{\dd \avg{x(\tau)A(0)}}{\dd \tau} 
\,\,.
\end{equation}
In single-particle tracking experiments the observable $A$ is the
tracer's position $x(t)$ and
$\avg{A(t)x(t')}=\avg{x(t)x(t')}=C_x(t,t')$ is the position
auto-correlation function, which, in equilibrium, is a function of
$t-t'$ only, $C_x(t,t')=C_x(t-t')$. The FDT in Eq.~\eqref{eq:fdt} has
the equivalent Fourier formulation
\begin{equation}\label{eq:fdtw}
\kb T = \f{-\omega \hat{C}_x(\omega)}{2 \hat{\chi}''(\omega)}
\,\,,
\end{equation}
where $\hat{\chi}''$ denotes the imaginary part of the response
Fourier transform (and where our convention for the Fourier transform
is
$\hat{f}(\omega)=\int_{-\infty}^\infty\text{d}t\;\text{e}^{-i\omega
  t }f(t)$). Alternatively, the FDT can be stated in the Laplace
domain in terms of the mean square displacement (MSD) $\avg{\Delta
  x^2}(t)=2(C_x(0)-C_x(t))$ as:
\begin{equation}\label{eq:fdtl}
\kb T = \f{\s \DXtilde(\s)}{2 \tilde{\chi}(\s)}
\,\,.  
\end{equation}
where the Laplace transform is defined by
$\tilde{f}(\s)=\int_0^\infty\text{d}t\; \text{e}^{-\s t}f(t)$.

In systems with a very small Reynolds number such as living cells,
that is when inertial effects are negligible---which includes the
Basset force---the response function is simply related to the Stokes
memory kernel in the Laplace domain by
$\tilde{\chi}(\s)=1/(\s\tilde{\g}(\s))$. The FDT is then usually
stated in terms of the complex modulus $\GG^*(\s)=\s\tilde{\eta}(\s)$
as~\cite{Lau,mason00,Mason}:
\begin{equation}
\DXtilde(\s) = \f{\kb T}{3\pi a\s \GG^*(\s)}
\,\,.
\end{equation}

\section{Noise correlations in equilibrium}

Our goal is now to explicitly derive the expression of the thermal
noise correlations $\avg{\xi(t)\xi(t')}=\sigma(t-t')$, as imposed by
the FDT in the presence of inertial effects.  By definition, the 
function $\sigma$ is even, $\sigma(t)=\sigma(-t)$. Here we follow the approach presented
in \cite{Maes,Bohec}. Since the thermal noise has Gaussian statistics,
the probability weight $\cP$ associated with a given realization of
the thermal noise is $\cP\brt{\xi}\propto \ee^{-\cA\brt{\xi}}$, where
$\cA\brt{\xi}= \frac12 \iint\limits_{t_0}^\infty \dd t_1\dd t_2
\xi(t_1)\Gamma(t_1-t_2)\xi(t_2)$. The expression of $\xi$ in this
formula is determined by the tracer's dynamics in Eq.~\eqref{eq:dyn},
and the symmetric function $\Gamma$ is related to the thermal noise
correlations by $\int\limits_{t_0}^\infty \dd t_1
\sigma(t-t_1)\Gamma(t_1-t')=\delta(t-t')$. The application of the
external {perturbation} $\fp$ results in a variation $\delta\cA$ of $\cA$, so
that the response function is expressed as:
\begin{equation}\label{eq:resp}
\chi(t,t') = -\avg{A(t) \left.\f{\delta\cA}{\delta \fp(t')} \right|_{\fp=0}}
\,\,.
\end{equation}
Substituting $\xi$ from Eq.~\eqref{eq:dyn} into $\cA\brt{\xi}$ and
calculating the functional derivative in Eq.~\eqref{eq:resp} yields 
the expression of the response function $\chi=\ci+\ce+\cs+\cb$, with
four contributions:
\begin{subequations}\label{eq:respb}
\begin{eqnarray}
\ci(t,t') &=& \mf \int\limits_{t_0}^\infty \dd t_1 \Gamma(t_1-t') \avg{\ddot x(t_1) A(t)}
\,\,,
\\
\ce(t,t') &=& - \int\limits_{t_0}^\infty \dd t_1 \Gamma(t_1-t') \avg{\Fe(t_1) A(t)}
\,\,,
\\
\label{eq:respS}
\cs(t,t') &=& \iint\limits_{t_0}^\infty \dd t_1\dd t_2 \Gamma(t_1-t') \g(t_1-t_2) \avg{\dot x(t_2) A(t)}
\,\,,
\nonumber
\\
\\
\label{eq:respB}
\cb(t,t') &=& \iint\limits_{t_0}^\infty \dd t_1 \dd t_2 \Gamma(t_1-t') \db(t_1-t_2) \avg{\ddot x(t_2) A(t)}
\,\,,
\nonumber
\\
\end{eqnarray}
\end{subequations}
where $\mf=m+m_\text{f}/2$. In order to compare this prediction with
the FDT, we focus on the regime where the system reaches an
equilibrium state, namely when the dynamics does no longer depend on
initial conditions, by setting $t_0\to-\infty$.  In this regime, the
correlation functions are time--translational invariant, and the
response functions depends only on the time lag $\tau=t-t'$.  We split
the difference $\chi(\tau)-\chi(-\tau)$ into four contributions
corresponding to the functions defined in Eq.~\eqref{eq:respb}. The
first contribution is expressed as:
\begin{eqnarray}
\ci(\tau)-\ci(-\tau) &=& \mf \int\limits_{-\infty}^\infty \dd t_1 \big[ \Gamma(t_1-\tau) \avg{\ddot x(t_1) A(0)}
\nonumber
\\
& &-\Gamma(t_1+\tau) \avg{\ddot x(t_1) A(0)} \big]
\,\,.
\end{eqnarray}
We perform the change of variable $t_1\to -t_1$ in the second
integral.  {The key trademark of equilibrium that we now make use of } is time reversibility, which implies, $\avg{\ddot
x(t_1) A(0)} = \avg{\ddot x(-t_1) A(t)}$.  Given that $\Gamma$ is
even, it follows $\ci(\tau)=\ci(-\tau)$, and we show similarly that
$\ce(\tau)=\ce(-\tau)$. We perform the changes of variable
$t_1\to-t_1$ and $t_2\to-t_2$ in the expression of $\cs$ and $\cb$,
and given the parity of the observables in the correlation functions
of Eqs.~\eqref{eq:respS} and~\eqref{eq:respB}, we deduce:
\begin{subequations}
\begin{eqnarray}
\cs(\tau)-\cs(-\tau) &=& \iint\limits_{-\infty}^\infty \dd t_1  \dd t_2 \Gamma(t_1-\tau)\avg{\dot x(t_2) A(0)} 
\nonumber
\\
& &\times\pnt{\g(t_1-t_2) + \g(t_2-t_1)}
\,\,,
\\
\cb(\tau)-\cb(-\tau) &=& \iint\limits_{-\infty}^\infty \dd t_1 \dd t_2 \Gamma(t_1-\tau) \avg{\ddot x(t_2) A(0)}
\nonumber
\\
& &\times\pnt{\db(t_1-t_2) - \db(t_2-t_1)} 
\,\,.
\end{eqnarray}
\end{subequations}
As a result, we finally obtain the expression of the difference
$\chi(\tau)-\chi(-\tau)$ in terms of the kernels appearing in the
generalized Stokes force and the Basset force. For an equilibrium
process, this expression should be identical to the prediction of the
FDT in Eq.~\eqref{eq:fdt}, which enforces that
\begin{eqnarray}\label{eq:fdtb}
&\beta& \avg{\dot x(\tau) A(0)}
\nonumber
\\
 &=& \iint\limits_{-\infty}^\infty \dd t_1 \dd t_2 \Gamma(t_1-\tau)
\nonumber
\\
& &\times\big[ \pnt{\g(t_1-t_2) + \g(t_2-t_1)} \avg{\dot x(t_2) A(0)}
\nonumber
\\
& &+\pnt{\db(t_1-t_2) - \db(t_2-t_1)} \avg{\ddot x(t_2) A(0)} \big]
\,\,.
\end{eqnarray}
This relation is independent of the parity of the observable we
consider for the response function. In the case where a more general
perturbation force $\fp(x)=-a_\text{\tiny P}(t) U'(x(t))$ is applied
to the tracer, it is also possible to define the response function
with respect to the parameter $a_\text{\tiny P}$:
\begin{equation}
\chi_\text{\tiny U}(t,t') = \left.\f{\delta \avg{A(t)}}{\delta a_\text{\tiny P}(t')}\right|_{a_\text{\tiny P}=0}
\,\,.
\end{equation}
We then recover the standard FDT (analogous to Eq.~(\ref{eq:fdtg})):
\begin{equation}
\chi_\text{\tiny U}(t-t')=\beta \frac{\p  \avg{A(t) U(x(t'))}}{\p t'} \Theta(t-t')\,\,.
\end{equation}
In the Fourier domain, and since the Fourier transform of thermal
correlations is related to $\hat{\Gamma}$ as:
$\hat{\sigma}(\omega)=1/\hat{\Gamma}(\omega)$, we obtain from
Eq.~\eqref{eq:fdtb}
\begin{equation}
\hat{\sigma} (\omega) = 2\kb T \pnt{ \hat{\g}'(\omega) - \omega \hat{\zeta}_\text{\tiny B}''(\omega) }
\,\,,
\end{equation}
where $\hat{\g}'$ and $\hat{\zeta}_\text{\tiny B}''$ denote the real
part of the $\g$ Fourier transform and the imaginary part of the $\db$
Fourier transform, respectively. Hence, we deduce the thermal noise
correlations read:
\begin{equation}
\label{eq:sigma_t}
\avg{\xi(t)\xi(t')} = \kb T \brt{ \g\pnt{\abs{t-t'}} + \f{\dd
    \zeta_\text{\tiny B}}{\dd t}\pnt{\abs{t-t'}} } \,\,.
\end{equation}
This result can be decoded as an effective visco--elastic memory
kernel $\gamma^*=\gamma + \dd \db/\dd t$, which could have been
guessed by integrating by parts the Basset memory term.  In that case
however integration by parts involves a $\db(0)$ term which at best is
not well defined, while our derivation encompasses this problem by
using an anti--symmetric function $\db(t)-\db(-t)$. Note that
$m_\text{f}$ does not appear in this expression, so that only the
terms with memory kernels in the Basset force and the generalized
Stokes force contribute to the dissipation of the tracer with the heat
bath as expressed by the FDT. This is in fully consistent with the free-paticle situation considered by Felderhof~\cite{:/content/aip/journal/jcp/131/16/10.1063/1.3258343} (his Eq. (2.10)) or by Indei {\it et al.}~\cite{Indei12b} (their Eqs. (64) and (65)). In the Laplace domain, the thermal
correlation function is expressed as:
\begin{eqnarray}
\label{eq:sigma_s}
\avg{\tilde{\xi}(\s)\tilde{\xi}(\s')} = \kb T \brt{
  \f{\tilde{\g}(\s) +\tilde{\g}(\s')}{\s+\s'} +
  \f{\s\tilde{\zeta}_\text{\tiny B}(\s)
    +\s'\tilde{\zeta}_\text{\tiny B}(\s')}{\s+\s'} }
\,\,.
\nonumber
\\
\end{eqnarray}

The equipartition theorem represents an alternative method to
characterize equilibrium properties. It relates the initial value of
the velocity autocorrelation function $C_v(t-t')=\avg{\dot x(t)\dot
x(t')}$ to the bath temperature as: $C_v(0)=\kb T/m$.  By using the
FDT prediction in Eq.~\eqref{eq:fdtl}, and given the velocity
autocorrelation function is simply related to the MSD in the Laplace
domain as: $\tilde{C}_v(\s)= \frac12 \s^2\DXtilde(\s)$, we deduce:
$\tilde{C}_v(\s)=\kb T\s\tilde{G}(\s)$, where $G$ denotes the
``usual'' response function
\cite{Desposito09,Grebenkov11}. Considering the dynamics described by
Eq.~\eqref{eq:dyn} with an external force $F_\text{ext}=-k x$, we
compute the response function in the Laplace domain, and we use
Eq.~\eqref{eq:kernw} to obtain:
\begin{equation}
\tilde{G}(\s) = \f{1}{\s^2\mf + 3 \s^{3/2}  \sqrt{m_\text{f}\tilde{\g}(\s)/2}+ \s\tilde{\g}(\s)+k}
\,\,.
\end{equation}
From the initial value theorem, we finally deduce:
\begin{eqnarray}\label{eq:EP}
\f{C_v(0)}{\kb T}= \lim_{\s\to\infty}
\f{1}{\mf +  3\sqrt{m_\text{f}\tilde{\g}(\s)/(2\s)}+
  \tilde{\g}(\s)/\s +k/s^2} \,\,.
\nonumber
\\
\end{eqnarray}
As discussed in Sec.~\ref{sec:mod}, the Laplace transform of the
Stokes memory kernel in the high frequency regime behaves like
$\s^{\alp-1}$, where $\alp<2$, so that:
$\tilde{\g}(\s)/\s\underset{\s\to\infty}{\longrightarrow}0$. It
follows that the initial value of the velocity autocorrelation
function $\avg{\dot x^2}=\kb T/\mf$ is different from the ``usual''
equipartition theorem prediction, as already noticed
in~\cite{Komura93}. Earlier works on this subject, like those of
Widom~\cite{PhysRevA.3.1394} or
Case~\cite{:/content/aip/journal/pof1/14/10/10.1063/1.1693298} used to
determine correlation functions by assuming the ``usual''
equipartition, leading to slightly wrong results. Here we show that
using the FDT as starting point avoids such issues. We also note that
this result remains the same if we consider a constant value for the
viscosity coefficient in the expression of the Basset force memory
kernel.  When an arbitrary external force $F_\text{ext}$ is applied to
the tracer, the initial value of the velocity autocorrelation function
satisfies Eq.~\eqref{eq:EP} under the modification $k\to
k(s)=-\tilde{C}_\text{ext}(s)/\tilde{C}_x(s)$, where
$C_\text{ext}(t)=\avg{x(t)F_\text{ext}(0)}$. This roughly means that
in the $s \to \infty$ limit $k$ can be replaced by $-\avg{x
  F_\text{ext}}_\text{eq}/\avg{x^2}_\text{eq}=\kb
T/\avg{x^2}_\text{eq}$.  Given the process defined in
Eq.~\eqref{eq:dyn} has a Gaussian statistics, an experimental method
to verify the validity of this result lies in measuring the stationary
distribution of the tracer's velocity~\cite{Raizen14}, for which the
variance should equal the initial value of the velocity
autocorrelation function. For an overdamped system in the absence of
external force, the condition $\avg{\Delta x^2}(0)=0$ associated with
the FDT prediction in Eq.~\eqref{eq:fdtl} imposes $\alp$ is positive,
meaning the Stokes kernel necessarily diverges in the short time limit
for such a system.

In summary, we have revised some equilibrium properties of generalized
Langevin equation with hydrodynamic interactions.  Under the
fluctuation-dissipation theorem, the memory kernels $\g(t)$ and
$\db(t)$ of generalized Stokes and Basset forces have been related to
the noise correlation function $\avg{\xi(t)\xi(t')}$ according to
Eq.~\eqref{eq:sigma_t}.  The derivation is valid in both Fourier and
Laplace domains.  This relation allows one to refine phenomenological
models that are used for the analysis and interpretation of
single-particle tracking experiments in complex visco-elastic media,
notably in living cells.  In particular, we showed that the noise
correlation function in \cite{Grebenkov13,Grebenkov14} should not
contain the term $m_{\text f} s/2$ which came from a naive extension
of the fluctuation-dissipation theorem to the Basset force (since this
term could alter tracer's dynamics only at very short time scales, its
presence does not affect the results reported in
\cite{Grebenkov13,Grebenkov14}).  Note also that relation
\eqref{eq:kernw} between the memory kernels of the generalized Stokes
and Basset forces allows one to reduce the number of model parameters
in \cite{Grebenkov13,Grebenkov14} yielding potentially more robust
fits.  Future optical tweezers single-particle tracking experiments at
short time scales can further clarify hydrodynamic interactions
between the tracer, the solvent, and semi-flexible polymers such as,
e.g., actin filaments.

\bibliographystyle{elsarticle-num}
\bibliography{fodor_revised} 

\end{document}